\documentstyle[12pt,bezier]{article}

              \def\be{\begin{equation}}
              \def\ee{\end{equation}}
              \def\bea{\begin{eqnarray}}
              \def\eea{\end{eqnarray}}
        \def\ba{$$\begin{array}}
        \def\ea{\end{array}$$}
        \def\d{\Delta}

\begin{document}
\begin{center}
{\Large{\bf{ Phase diagram of the anti-ferromagnetic {\bf xxz} model  \\
in the presence of an external magnetic field}}}
\vskip 1.5 cm
{{{\bf A. Langari}\footnote{E-mail : \it {langari@netware2.ipm.ac.ir}} }}   \\
\vskip 1cm
{\it \small { Institute for Studies in Theoretical Physics
and Mathematics
\\ P.O.Box: 19395-5531, Tehran, Iran. \\  Department of Physics,
Sharif University of Technology\\ P.O.Box:11365-9161, Tehran, Iran  }}
\end{center}
\vskip 1cm
\begin{abstract}
The anisotropic $s=\frac{1}{2}$ anti-ferromagnetic Heisenberg
chain in the presence of an external magnetic field is studied by using the 
standard quantum renormalization group. We obtain the critical 
line of the transition from partially magnetized (PM) phase 
to the saturated ferromagnetic (SFM)
phase. 
The crossover exponent between the PM phase 
and anti-ferromagnetic Ising (AFI) phase is evaluated. Our
results show that the anisotropy($\d$) term is relevant and causes
crossover. These results indicate that the standard RG approach
yields fairly good
values for the critical points and their exponents.
The magnetization curve, correlation functions and the ground state
energy per site are obtained and compared with the known exact results.
\end{abstract}

PACS numbers : 05.50.+q,75.10.-b,75.10.Jm \\
Keywords : Quantum renormalization group, Spin systems, Heisenberg model.

\newpage
\def\bra#1{{\langle #1 \vert}}
\def\ket#1{{\vert #1 \rangle}}
\def\x#1{{s_{#1}^{x}}}
\def\z#1{{s_{#1}^{z}}}
\def\y#1{{s_{#1}^{y}}}
\def\a{\alpha}
\def\b{\beta}

\section{Introduction}
Systems near criticality are usually characterized by fluctuations over
many length scales.
At the critical point itself, fluctuations exist over
all scales. At moderate temperatures 
quantum fluctuations are usually suppressed compared with the
thermal fluctuations. However if the temperature is near zero, the quantum
fluctuations especially in the low-lying states dominate the thermal ones
and strongly influence the critical behaviour of the systems. The study
of the ground state and its energy is thus of centeral importance for
understanding the critical behaviour of such systems.

The technique of Renormalization
Group (RG) has been so devised to deal with these multi-scale 
problems\cite{r1,r2,r3,r4}. In the momentum space RG which is suitable
for studying continuous systems one iteratively integrates out the small scale
fluctuations and renormalizes the Hamiltonian. In the real space RG, which is
usually performed on lattice systems with discrete variables (i.e. quantum
spin chains), one divides the lattice into blocks which are treated as the
sites of the new lattice\cite{r5}. The Hamiltonian is divided into intra-block
and inter-block parts, the later being exactly diagonalized, and a number of
low lying energy eigenstates are kept to project the full Hamiltonian onto the
new lattice. The accuracy
of the method is determined by the number of states kept and specially is
sensitive to the boundary conditions\cite{r6,r7,r8} which is considered
for the block Hamiltonian.
The detailed form of this projection in fact differentiates various
versions of the real space RG, ranging from the standard RG to the recent
Density Matrix RG (DMRG)\cite{r9}. Each of these versions have their own
advantage and disadvantages.

The Ising model in a transverse field (ITF) and anisotropic XY model 
in a transverse field (AXYTF) have been studied in \cite{r10,r11} 
using both standard RG and DMRG methods. There, it has been concluded
that DMRG gives accurate results for the ground state energy and correlation functions
in both models, but the standard RG method where the number of states kept are not
few can give better results in determining the location of critical points
and critical exponents. In this direction we have been motivated to study a more
general model, the anisotropic Heisenberg model in the presence of an external magnetic
field ({\bf xxz+h}) by using the standard RG method to compare its results
with the known exact ones.
This RG study allows us to have analytic RG equations which gives a better
understanding of the behaviour of the real space RG method at 
the critical points.
We have studied the {\bf xxz+h} model because of its richness in 
the phase-diagram
where there are different critical behaviour.
In this study we have succeeded
to obtain the critical line between the partially magnetized (PM)  
and saturated ferromagnetic (SFM) phases, to a good accuracy, compared with 
the known results\cite{r12}.
We have also derived the crossover between the PM phase (small anisotropy
$-1<\Delta<1$) and the anti-ferromagnetic Ising (AFI) phase (large anisotropy
$\Delta>>1$) and calculated its exponent ($\phi>0$) which verifies the
relevance of anisotropy to the crossover phenomena. These 
results which come out
of an RSRG by keeping only two states in each block confirm that RSRG
is a good candidate to study at least the qualitative behaviour of the quantum
lattice systems in the quantum critical region.

In this letter we have studied the {\bf xxz+h} model by RSRG method where
the block length is three ($n_B=3$). In the next section we will introduce the
{\bf xxz+h} model and discuss its critical behaviour as derived by other
methods. In section 3 we will discuss about different types of constructing
RG--equations and obtain the analytic RG--equations for this model.
Using these equations we will describe the phase-diagram of this model. We will
discuss the critical behaviour of the {\bf xxz+h} model by RG--equations in
section 4. In section 5 we will compare some of the results i.e.
ground state energy and correlation functions with the known exact results.
The paper ends with a conclusion.
\section{The Model}
The anisotropic spin $\frac{1}{2}$ anti-ferromagnetic Heisenberg model, 
or {\bf xxz} chain
is one of the most studied quantum spin systems in statistical mechanics.
It is also a classic example of one dimensional integrable quantum 
spin systems\cite{r12,r13,r135}. 
The {\bf xxz} chain gives us the first example of a critical line
with critical exponents varying continuously with the anisotropy.
The model has also been studied by using the conformal invariance idea\cite{r14},
where the critical fluctuations along the critical line are
governed by a conformal field theory with central charge c=1. The RG
study of {\bf xxz} model has been performed by Rabin\cite{r15}. At $\Delta=0$,
even in the presence of magnetic field, the model can be mapped 
by Jordan--Wigner transformation to
free--fermions\cite{r16}, which is exactly solvable.
This system has also been studied within the RG formalism\cite{r11}.

The Hamiltonian in the presence of an external magnetic field is
\be
H(J,\Delta,h)=J \sum_{i=1}^{N}(\x{i}\x{i+1}+\y{i}\y{i+1}
+\d \z{i}\z{i+1}+h \z{i})
\ee
where $J>0$, $\d$ is the anisotropy parameter, which
in the anti-ferromagnetic region is taken
to be greater than or equal to $-1$, and $h>0$ is the strength of 
the external magnetic field.
The effect of a uniform magnetic field in the phase diagram of the {\bf xxz}
chain is to extend the critical phase over a finite region which is 
partially magnetized and delimited by a critical line where the chain
becomes saturated ferromagnetic\cite{r12,r14}. 
Uniform external magnetic field does not
destroy the exact integrability of the quantum chain but the coupled
integral equations for the spectral parameter do not have closed 
analytic solutions. Then the only results are numerical
or perturbative ones\cite{r12,r135,r14}.

The Hamiltonian $H(J,\d,h)$ is related by a canonical transformation
$U=e^{i\pi \sum_{j=1}^N j \z{j}}$ to $H(J,-\d,h)$, i.e.
$UH(J,\d,h)U^{-1}=-H(J,-\d,h)$. This gives a relation between the 
antiferromagntic($J>0$) and ferromagnetic($J<0$) cases. 
At $\d=1$ and $h=0$ the Hamiltonian exhibits an
su(2) symmetry. For $\d\neq1$,
it exhibits a quantum symmetry $su_q(2)$\cite{r165}.
If $h\neq0$, the only symmetry is U(1). Let us now begin the RG study of
the {\bf xxz+h} model.
\section{Renormalization Group Equations}
The implementation of RSRG is based on two important points, 
the size of blocks
and the number of states kept in each step of RG. Both of them would have
significant effects on the RG flow. Here we choose a 3-sites block($n_B=3$)
for the renormalization process. In this case the two lowest energy states
of the block Hamiltonian preserve the symmetries of the Hamiltonian 
and lead to a self similar Hamiltonian.
Moreover at large $\d$ and $h$ the level crossing of the 
ground state in the block
occurs at a coupling constant which is exactly its critical value (this will be
explained latter, see also appendix). Finally taking larger blocks, renders 
an analytic RG--equations, difficult to obtain.

After dividing the whole chain into 3-sites blocks, the first step of RSRG
is to divide the Hamiltonian into two parts, the intra--block Hamiltonian $(H^B)$ and the
inter--block Hamiltonian 
($H^{BB}$). There are several choices for doing this decomposition.
In our prescription
we choose the decomposition which is sketched in fig.1. In this case
the block Hamiltonian is 
\be
H^B_{\mu}=J[\x{1}\x{2}+\x{2}\x{3}+\y{1}\y{2}+\y{2}\y{3}
+ \d(\z{1}\z{2}+\z{2}\z{3})+h(\z{1}+\z{2}+\z{3})],
\ee
where $\mu$ represents the block number i.e. $H^B=\sum_{\mu=1}^{N/3} H^B_{\mu}$.
The block Hamiltonian is diagonalized exactly and 
then the two lowest lying states are
kept to span the truncated (or effective) Hilbert space.
Thus the embedding operator(T) is constructed to be
\be
T=\ket{\alpha}\bra{+} + \ket{\beta}\bra{-}
\ee
where $\ket{\alpha}$ and $\ket{\beta}$ represent the two low-lying eigenstates
of $H^B$ and $\ket{+} , \ket{-}$ are the renamed base kets for the effective
Hilbert space.

There is a level crossing at $h=h_0$ (eq.(\ref{ho})) 
in the spectrum of $H^B_{\mu}$,
where the ground state
changes from the $S^z=\frac{-1}{2}$ state to the $S^z=\frac{-3}{2}$ state
($S^z=\z{1}+\z{2}+\z{3}$). Note that $h_0$ comes from the finite size effects
of a 3-sites block and reaches the critical 
value of external magnetic field ($h_c$)
by increasing the block sizes ($n_B \rightarrow \infty$). In that case the
ground state changes from a PM state ($m\neq0$) to a SFM state
($|m|=0.5$). Thus in the absence of magnetic field, the ground state of 
the block  Hamiltonian is a spin $\frac{1}{2}$ doublet.
As $h$ is turned on weakly, we enter in a Zeeman regime and the doublet 
splits into two states. This is true as long as $h<h_0$.
For strong magnetic fields, we are in a regime in which the ground state
is a singlet with all spins down. This should correspond to $h>h_0$.
The results of this computation are as follows :
\vspace{0.5cm}
for $h<h_0$ :
\bea
\ket{\alpha}&=&b\ket{++-}+a\ket{+-+}+b\ket{-++} ,  \\ \nonumber
\ket{\beta}&=&-b\ket{--+}-a\ket{-+-}-b\ket{+--}  ,
\eea
where
\be\label{ho}
h_0(\d)= \frac{3 \d+\sqrt{\d^2 +8}}{4},  
\ee
$$
a=\frac{2x+2}{\sqrt{6+12x^2}} \hspace{1cm},
\hspace{1cm}b=\frac{2x-1}{\sqrt{6+12x^2}}   
$$
and
$$
x=\frac{2(\d-1)}{8+\d+3\sqrt{\d^2 +8}}.  
$$
for $h>h_0$ :
\bea
\ket{\alpha}&=&-b\ket{--+}-a\ket{-+-}-b\ket{+--} ,  \\ \nonumber 
\ket{\beta}&=&\ket{---}.
\eea
Having the embedding operator at hand, the operators (observables) 
are renormalized as
\be
O'=T^{\dagger}OT.
\ee
By using the above equation one can obtain the renormalization of operators.
Thus for $h<h_0$ we obtain the following relations : ($h<h_0$)
\bea\label{r1} 
T^{\dagger} \x{1(3)} T = -2ab s'^x \hspace{0.5cm}&;&\hspace{0.5cm}
T^{\dagger} \x{2} T = -2b^2 s'^x  \\ \nonumber
T^{\dagger} \y{1(3)} T = 2ab s'^y \hspace{0.5cm}&;&\hspace{0.5cm}
T^{\dagger} \y{2} T = 2b^2 s'^y  \\ \nonumber
T^{\dagger} \z{1(3)} T = a^2 s'^z \hspace{0.5cm}&;&\hspace{0.5cm}
T^{\dagger} \z{2} T = (1-2a^2) s'^z  
\eea
We find the same renormalization for $s_1$ and $s_3$ because of the symmetry
in site $1$ and $3$ in the block (1---2---3). We will obtain 
similar relations for $h>h_0$ :
\bea\label{r2}
T^{\dagger} \x{1(3)} T = -b s'^x \hspace{0.5cm}&;&\hspace{0.5cm}
T^{\dagger} \x{2} T = -a s'^x  \\ \nonumber
T^{\dagger} \y{1(3)} T = b s'^y \hspace{0.5cm}&;&\hspace{0.5cm}
T^{\dagger} \y{2} T = a s'^y  \\ \nonumber
T^{\dagger} \z{1(3)} T = - \frac{a^2+1}{4} I + \frac{1-a^2}{2} s'^z \hspace{0.5cm}&;&\hspace{0.5cm}
T^{\dagger} \z{2} T = \frac{a^2-1}{2} I + a^2 s'^z  .
\eea
In the above equations $s'^{\alpha}$ is the effective operator in the effective
Hilbert space of the block (new sites in the renormalized chain). By considering 
the interaction between blocks and using the above equations (\ref{r1}),(\ref{r2})
we will obtain the renormalization of coupling constants in the Hamiltonian :
\be\label{rg1}
|h|<h_0(\d) \;\;{\bf :}\hspace{4cm} \left\{ \begin{array}{cc}
J'=4 a^2 b^2 J,  \\
\d'=\frac{a^2}{4 b^2} \d,  \\
h'=\frac{1}{4 a^2 b^2} h,  
\end{array}\right.
\ee

\be\label{rg2}
|h|>h_0(\d) \;\;{\bf :}\hspace{2cm} \left\{ \begin{array}{cc}
J'=b^2 J,  \\
\d'=b^2 \d,  \\
h'=\frac{1}{b^2}(|h|-h_0+\frac{a^4-1}{4} \d) sgn(h).
\end{array}\right.
\ee
The above RG--equations show that the renormalized Hamiltonian 
is of the same form as the original one. 
The critical behaviour which can be obtained from these equations
will be discussed in the next section.

Let us consider an extreme case where
both $h$ and $\d$ go to infinity and $J$ goes to zero 
such that $J\d$, $Jh$ and $\frac{h}{\d}$ remain finite. 
In this case the Hamiltonian reduces
to a simple anti-ferromagnetic Ising (AFI) model 
in the presence of an external magnetic field, which shows a first
order transition from a classical anti-ferromagnetic(Neel) ordered phase ($m=0 ,
sm=\frac{1}{N} \sum_{i} (-1)^i \z{i}=0.5$) to a saturated ferromagnetic phase (SFM)
($|m|=0.5, sm=0$). We can write this Hamiltonian as
\be\label{afi}
H_{AFI}=k \sum_{i}(\z{i}\z{i+1}+g\z{i})
\ee
where
\be\label{aficoupl}
k=J\d>0 \hspace{1cm},\hspace{1cm} g=\frac{h}{\d}   \nonumber
\ee
At large $\d$ ($\d>>1$) we have $h_0\simeq\d\;\;, a\rightarrow1\;\; and \;\;b\rightarrow0\;$.
Then the RG--equations reduces to the following equations.\\
 
\be\label{caf1}
|h|<\d \;\;{\bf :}\hspace{4cm} \left\{ \begin{array}{cc}
J'=4 b^2 J,  \\
\d'=\frac{1}{4 b^2} \d,  \\
h'=\frac{1}{4 b^2} h,   
\end{array}\right.
\ee
  
\be\label{caf2}
|h|>\d \;\;{\bf :}\hspace{3cm}\left\{ \begin{array}{cc}
J'=b^2 J,  \\
\d'=b^2 \d,  \\
h'=\frac{1}{b^2}(|h|-\d) sgn(h).
\end{array}\right.
\ee
These RG--equations give exactly the critical point $g_c = 1$
and the ground state energy of the AFI model(see appendix), which will be discussed in the next section.
\section{Critical Behaviour}
In this section we analyze the RG--equations and its critical behaviour.
The phase diagram of the obtained RG--flow(eqs.(\ref{rg1},\ref{rg2})) is depicted 
in fig.2. This phase diagram
consists of three different phases, partially magnetized (PM),
classical anti-ferromagnetic(AFI)
and saturated ferromagnetic (SFM) phases.

There are five fixed points in the phase diagram, \\
({\it i})  XX represents a spin $\frac{1}{2}$ {\bf xx} model 
without external field,  \\
({\it ii}) XXTF is the critical point of the {\bf xx} model in the
presence of a transverse field, \\
({\it iii}) IAFH represents the critical
point of the {\bf xxz} model in the absence of an external field, \\
({\it iv}) AFI represents a classical anti-ferromagnetic
Ising model with a long-range Neel order,\\ 
({\it v}) SFM represents a saturated ferromagnetic phase where all spins align
in the direction of the external field.

In the SFM phase the RG--flow has a well defined behaviour and goes to the SFM
fixed point for any value of $h>h_c(\d)$. But when $h<h_c(\d)$ and 
$-1\leq\d\leq1$ the RG--flow represent a massless phase in which 
$J^{(n)}\rightarrow0$ and $\d^{(n)}\rightarrow0$
in the limit $n\rightarrow\infty$ 
($n$ is the number of RG steps). 
The RG flow in the PM phase has a cyclic nature. Since it reflects a 
sequence of level crossings between states with different values
of the total $S^z$ induced by varying the magnetic field.
The recurrence of this level crossing in the process of RG leads to the
oscillatory behaviour of the RG flow. If we imagine of a 3--dimensional
RG flow, its projection onto the $h-\d$ plane will have some closed 
paths. However if the initial point is in the PM(or SFM) phase,
it will go to the XX(or SFM) fixed point finally.
Therefore we conclude that the RG flow in the PM phase can be sketched
as in fig.2.
At the end of this region on the $\d=0$ line there
exists a fixed point at $h_c=0.943$ which separates the PM and SFM phases.
The eigenvalues of RG--flow at this critical point are given in table-1, which
give a relevant direction on the $\d=0$ line and an irrelevant direction along
the critical line $h_c=0.943 \d+0.943$. This critical line is obtained from
the behaviour of the correlation functions. We have calculated
$\bra{0}\z{i}\z{i+1}\ket{0}$ and plotted it versus $h$ (fig.5) which shows the
entrance to the SFM phase for $h>h_c(\d)$ in the $-1\leq\d\leq1$ region.
It is an interesting result which is obtained by a 3-sites block RG and
can be compared with the exact result $h_c=\d+1\;\;$\cite{r12}.
Although the obtained critical value for $h_c(\d)$ has a slight difference
in the coefficients but preserves the linear form of the critical line.
Our data for the ground state energy ($e_o$) show that $h_c(\d)$ 
represents a critical
line of 2nd order transition in which $e_o$ and $\frac{d e_o}{d h}$ are
continuous at $h_c(\d)$ and is confirmed by the analytic results for $e_o(h)$
at $\d=0\;$\cite{r16}. Our results along the
$\d=0$ and $h=0$ lines recover the results of 
Drzewinski\cite{r11} and Rabin\cite{r15} respectively.

For the XXTF  fixed point we have calculated the critical exponents which
have been written in table-1. If $R(h)$ represents the 
renormalization of $h$ along the
$\d=0$ line the correlation length exponent ($\nu$) is given by
$\nu =\frac{ln(n_B)}{ln(R'(h^*))}$. The dynamical exponent ($z$) is
$z=\frac{ln(\frac{J}{J'})_{h^*}}{ln(n_B)}$. The critical exponent $\alpha$
connected with the specific heat is calculated from the hyper-scaling relation
$2-\alpha=d^* \nu\;\;$\cite{r17}, where $d^*=d+z$ 
($d$ is the spatial dimension). The critical exponent $\beta$, related
to the magnetization is given by $\beta=\frac{ln(\frac{m'}{m})}{ln(R'(h^*)}$.
These results show good agreement with the exact ones.

The other phase in the phase diagram is a classical anti-ferromagnetic phase. Let us first
look at the exact solution of this model with the Hamiltonian 
as in eq.(\ref{afi}).
By a simple argument we can find that the ground state is a Neel ordered
state whose energy per site is $\frac{E_o}{k N}=e_o=\frac{-1}{4}$ for $0\leq g \leq1$ 
and is a saturated
ferromagnetic state for $g\geq1$ where $e_o=\frac{1-2g}{4}$ (see appendix). 
These values for the ground
state energy($e_o$) show a discontinuity of $\frac{d e}{d g}$ 
at $g=1$, which means the phase
transition at this point is classified as a 1st  order transition.
By using the previous definitions for $g=\frac{h}{\d}$, the RG--equations
in (\ref{caf1}) and (\ref{caf2}) give a fixed point at $g=1$ 
in the limit $\d \rightarrow \infty$, which is equal to 
the exact critical point.
The RG--flow in (\ref{caf1}) and (\ref{caf2}) show a fixed line at $\d\rightarrow\infty$
for all $g<1$. This means that there is a unique ground state for any 
value of $0\leq g \leq1$ and the
distinction between two different $g$ values is only due to the excited states
of the Hamiltonian.

We have also calculated the ground state energy 
by using the RG--equations (\ref{caf1}) and (\ref{caf2}) in a
hierarchical way by accumulating the energies of the blocks (see appendix). 
The obtained
result is equal to the exact result for the ground state energy.
\be
e_o=\left\{ \begin{array}{cc}
\frac{-1}{4} \;\;\;\;\;\;\; 0\leq g \leq1 \\
\frac{1-2g}{4} \;\;\;\;\;\; g \geq 1
\end{array}\right.
\ee
Thus the limiting case of our RG--equations at $h,\d>>1$ exactly
describe the classical anti-ferromagnetic Ising model.

There is an interesting point in the phase-diagram. When $-1\leq\d\leq1$ and $h<h_c(\d)$,
the model represent a PM phase with $-0.5<m\leq0$ and undergoes a 2nd
order transition to the SFM phase($m=-0.5$) at $h_c(\d)$(m is continuous at the 
transition point). 
But at $h,\d \rightarrow\infty; \frac{h}{\d}<1$ when
the model represent a Neel ordered phase (AFI) with $m=0$ and $sm=0.5,$
it undergoes a 1st order transition to the SFM phase($m=-0.5$) at $h_c=\d$
(m is discontinuous at the transition point). We have
examined the continuity of the ground state energy and 
its derivatives up to the third order
numerically at $\d=1 ; h\neq0$. Therefore there is a continuous change in the
critical exponents by increasing $\d$ when $h<h_c$. This shows a crossover
between PM and AFI phases. This change in the 
universality class is due to the reduction in the number 
of components of the spin
operator (i.e. three component $\x{},\y{},\z{}$ in the PM phase and effectively
one component $\z{}$ in the AFI phase).
The crossover exponent
$\phi=\frac{y_{\d}}{y_h}$ has been calculated to be $0.63$ which verifies
that the coupling $\d$ is relevant and causes crossover
($\lambda_i=n^{y_i}_B$ ; where $\lambda_i$ is the eigenvalue at IAFH fixed point).

One can also retrieve the critical line $0<\d<1\;;\;h=0$ by RSRG using
the $su_q(2)$ symmetry\cite{r18} of the {\bf xxz} chain 
which differs only at the ends and is not important 
in the thermodynamic limit ($N\rightarrow\infty$).
However it is not suitable
for $\d>1$ case and also in the presence of the external field 
($h$) this RG prescription could not describe the critical line $h_c(\d)$.
\section{Energy and Correlation Functions}
In this section we describe some more results which have been obtained by RG--equations
in section-3. In fig.3-a we have plotted the ground state energy per site($e_o$) of the
XX($\d=0$) model versus external magnetic field($h$). We have also compared the RSRG results
with the exact one\cite{r16}. In the PM phase (fig.3-a) there is a discontinuity in 
$e_o$ which is due to level crossing(finite size effects) of 3-sites block.
This level crossing occurs as $h_0(\d=0)$ passes the value $0.707$.
Thus we have not considered
this  point as a critical point. Fig.3-a shows good agreement with the exact
results in the SFM phase($h>h_c$) with a slight difference in the PM phase. It
has been shown\cite{r6,r7,r8} that this difference is due to boundary conditions
in an isolated block which neglects the remaining part of the chain. We have
introduced a modified scheme to decrease this difference in the $h=0$ case 
\cite{r8}. By using the RG--equations of section-3 the ground state energy for
different values of $\d$ can be calculated which will show the same behaviour as
in fig.3. We have plotted the ground state energy for $\d=-1, -0.5, 0.5, 1$ 
cases in fig.3-b. The ground state energy at $\d=-1$ in fig.3-b confirms 
that the model represents the ferromagnetic Heisenberg model in the presence 
of an external magnetic field, where its ground state energy is proportional 
to the strength of the external field.

We have plotted in fig.4-a, the magnetization($m$) versus
external magnetic field($h$) for $\d=0$. It has been compared with the known exact result, which
shows good agreement qualitatively. The step form of the RSRG results in this
figure is due to the cyclic nature of the RG--equations in the PM phase.
This is related to the nature of the anti-ferromagnetic problem.
The magnetization curve reflects a continuous sequence of level-crossings
between states with different values of the total $S^z$
induced by varying the magnetic field. 
The variational ground state
which is obtained here is however owing to the highly degenerate energy
level structure.
The recurrence of this level crossing in the process of RG leads to the 
oscillatory behaviour of the RG flow.
This oscillation is trapped by a metastable state which leads to a jump
in the magnetization curve.
Fig.4-a confirms that for $h<h_c$ the model is PM where $m\neq0$ and reaches
the SFM phase ($m=-0.5$) at $h=h_c$. We have also plotted the magnetization
versus external field in fig.4-b for $\d=-0.5, 0.5$ and fig.4-c 
for $\d=1.1, 1.5$ which show a similar 
behaviour as in the $\d=0$ case, but the critical point where the model
becomes SFM is different.

One of the important quantities which can be used to show the critical
behaviour is the z--component of spin--spin correlation function. We have
plotted $\bra{0}\z{i}\z{i+1}\ket{0}$ versus external magnetic field($h$) in fig.5 for different
values of anisotropy parameter $\d$. Transition from the PM phase to the SFM
phase occurs at two steps. The first jump is due to level crossing at
$h_0(\d)$ which is not the critical point and does not terminate at the SFM 
phase. But the second jump which achieves the SFM phase
($\bra{0}\z{i}\z{i+1}\ket{0}=0.25$), corresponds 
to the critical point of the transition from the PM to the SFM phases. 
The location of critical point is at the same
point in which the magnetization curves (fig.4-a, fig.4-b and fig.4-c) reaches the saturated 
value ($m=-0.5$).
The critical point $h_c(\d)$ for different values
of $\d$ which has been obtained in fig.5 yields the linear relation
$h_c(\d)=0.943\d+0.943$. This result shows that RSRG is a good candidate
to describe the critical behaviour of quantum systems\cite{r11}. 
We have also calculated the magnetization(m) in the middle range $\d>1$
and found that the critical line $h_c(\d)=0.943\d+0.943$ is also valid
in this region. Since the RSRG method is an approximate scheme there is 
a small discrepancy between the obtained critical line and the exact 
one ($h_c=\d+1$). This error is related to the small isolated blocks which 
are considered in this method. In other words the quantum fluctuations 
of a highly correlated system in a large lattice can not be simulated 
by a small number of eigenkets of an isolated block. However as
$\d\rightarrow\infty$ and the model becomes classical one, exact results
can be obtained by RSRG method (see appendix). This describes the 
discrepancy in the limit $\d \rightarrow \infty$ of $h_c(\d)$ which is 
$g_c=\frac{h_c}{\d}=0.943+O(\frac{1}{\d})$ and $g_c=1$ which can be 
obtained by RSRG method in the limit $\d \rightarrow \infty$ of the 
initial model.
We have also calculated the z-component of spin--spin correlation function
in terms of distance. In fig.6 we have plotted $\bra{0}\z{i}\z{i+r}\ket{0}$
at $\d=1$ for different values of $h$ below 
and above its critical point($h_c=1.886$).
When $h=0$, the correlation length is small(in the order of lattice spacing)
and the correlation function goes rapidly to zero after a few lattice spacing.
As $h$ increases to its critical value the correlation function becomes
nonzero for long distances and shows exactly the SFM phase above the critical
point($h>h_c$). The same behaviour has also been observed for other values
of the anisotropy parameter $\d$.
\section{Conclusion}
We have considered the anisotropic anti-ferromagnetic Heisenberg 
chain in the presence of an external magnetic field by RSRG. 
We have sketched the
phase diagram of this model for $\d\geq-1$ and $h>0$ in fig.2, 
the phase diagram for 
$\d\geq-1$ and $h<0$ is the mirror image of the previous case. 
We have obtained three distinct 
phases in the phase diagram. The partially magnetized (PM) phase with 
$m\neq0$, the saturated ferromagnetic (SFM) phase with $m=-0.5$ and the
Neel ordered phase (AFI) where $m=0$. By computing the magnetization and the
z-component of spin-spin correlation function we have calculated the 
critical line $h_c(\d)=0.943\d+0.943$ between the PM and SFM phases
which causes a 2nd order transition. But at $\d\rightarrow\infty$ 
the transition from
AFI to SFM phases is 1st order at the critical point $g_c=1$. We have 
observed that for $h<h_c$ increasing
$\d$ causes a crossover between PM and AFI phases which changes 
the universality class of the model. 
The crossover exponent
has been calculated to be $\phi=0.63$ which confirms the relevance of 
the anisotropy $\d$ in the crossover phenomena.

By using the analytical RG--equations we have obtained the critical exponents
at the XXTF fixed point. Although the obtained critical exponents are not 
accurate compared with the exact results they show good agreement with them.
The ground state energy and correlation
functions calculated in the PM phase, show qualitatively good 
results, but some discrepancy due to the boundary conditions of the 
isolated block in the RG procedure is present. 
However all the results in the SFM phase are completely accurate,
because the ground state of the whole chain in this phase is a simple 
juxtaposition of the ground state of the isolated blocks and there is not 
any boundary condition effects for an isolated block as in the PM phase.
In the AFI fixed point when the model is a classical one,
the limiting form of our analytical RG--equations ($\d\rightarrow\infty$) 
give the exact 
results for the ground state energy and the critical point $g_c=1$.
We have shown that the critical line $h_c(\d)$   
is also valid in the middle range $\d>1$.
Finally we conclude that the standard quantum RG (RSRG) gives qualitatively
good results for the critical behaviour of the system. However the quantitative 
results for the location of the critical point and critical exponents is 
much better than its results for ground state energy and correlation functions
with respect to the known exact results.
\section{Acknowledgement}
I would like to express my deep gratitude to V. Karimipour and J. Davoodi
for valuable comments and discussions and M.A. Martin-Delgado
for a careful study of the manuscript and very useful comments.
Interesting conversations with
B.Davoodi, M.R.Ejtehadi, M. R. Rahimi-Tabar, K. Kaviani, R. Razmi 
and M. Abolfath are also acknowledged.

\section{Appendix :  Classical anti-ferromagnetic Ising model in an external field} 
\subsection {Exact ground state}
The classical anti-ferromagnetic Ising model in the presence of 
an external magnetic field is given by equation (\ref{afi})
$$H_{AFI}=k \sum_{i=1}^N (\z{i}\z{i+1}+g\z{i})$$
where $k=J\d>0$ and  $g=\frac{h}{\d}$ is the strength of external magnetic 
field. We assume periodic boundary condition $\z{N+1}=\z{1}$. Let
us write the Hamiltonian in terms of Pauli matrices,
\be
H_{AFI}=\frac{k}{4}[ \sum_{i=1}^N (\sigma^z_{i} \sigma^z_{i+1}-1)
+2g\sum_{i=1}^N \sigma^z_{i}+N].
\ee
The energy ($E$) of this chain is 
\be
E=\frac{k}{4}[-2n_f+2g n_s+ N],
\ee
where $n_s=n_p-n_m$, $n_p$ is the number of up spins, $n_m$ is the number of 
down spins ($n_p+n_m=N$) and $n_f$ is the number of boundary walls of flipped 
spins. The maximum value of $n_f$ is obtained by a Neel ordered state
($n_p=n_m=\frac{N}{2} , n_f=N$), but for an arbitrary value of $n_s$
it can be written as
\be   (n_f)_{max}=N-|n_s|.   \ee
By using the definition of $n_s$, $(n_f)_{max}$ can be written in the 
following form
\be
(n_f)_{max}=N-|N-2n_m|=\left\{ \begin{array}{cc}
2n_m \;\;\;\;\;\;\;\;\;\;\;\;\;\;\;\; n_m\leq\frac{N}{2} \\
2(N-n_m) \;\;\;\;\;\; n_m\geq\frac{N}{2}
\end{array}\right.
\ee
The minimum value of $E$ will be obtained if $n_s$ has its minimum value and
$n_f$ has its maximum value.
\be   E_o=min(E)=\frac{k}{4}[N+2g(N-2n_m)-2(n_f)_{max}] .   \ee
When $n_m\leq\frac{N}{2}$, we have $\frac{E_o}{k N}=e_o=\frac{-1}{4}$ 
which is greater than all the energies in the $n_m\geq\frac{N}{2}$ case.
Therefore we will investigate $E_o$ in the $n_m\geq\frac{N}{2}$ region 
by minimizing $E$ with respect to $n_m$, which is 
\be\label{star}
e_o=\left\{ \begin{array}{cc}
\frac{-1}{4} \;\;\;\;\;\;\;\hspace{1cm} g \leq1 \;\;\;(n_m=\frac{N}{2})\\
\frac{1-2g}{4} \;\;\;\;\;\;\hspace{1cm} g \geq1 \;\;\;(n_m=N)
\end{array}\right.
\ee
It is obvious from (\ref{star}) that for any value of $g\leq1$ the ground
state energy is due to a Neel ordered state ($n_m=\frac{N}{2} ,
n_f=N$) and for $g\geq1$ the ground state is a saturated ferromagnetic
state ($n_m=N , n_f=0$).

\subsection {Renormalization group equations}
At large $\d (\d\rightarrow\infty)$, the renormalization group equations
(\ref{caf1}) and (\ref{caf2}) can be written for 
$k=J\d>0$ and  $g=\frac{h}{\d}\geq0$ in the following form
\be\label{gl1}
g<1 \;\;{\bf :}\hspace{3cm} \left\{ \begin{array}{cc}
k'=k,  \\
g'=g,  \\
\end{array}\right.
\ee
\be\label{gg1}
g>1 \;\;{\bf :}\hspace{3cm}\left\{ \begin{array}{cc}
k'=b^4 k,  \\
g'=\frac{1}{b^4}(g-1).  \\
\end{array}\right.
\ee
The RG equations in (\ref{gl1}) give no running of coupling constants
and lead to a fixed line $0\leq g \leq 1$. As far as the ground state energy
is concerned,
there is no distinction between any arbitrary
value of $0\leq g \leq 1$. But when $g>1$ the only fixed point is $g^*=1$
which is at the end of the fixed line $0\leq g \leq 1$. Thus $g^*=1$ is the 
critical value of $g$ which separates the $g<1$ and $g>1$ phases.
To be more rigorous and define these phases we will calculate the 
magnetization ($m$) and staggered magnetization ($sm$) in these regions.
Equations (\ref{r1}) and (\ref{r2}) in the limit $\d\rightarrow\infty$
($a \rightarrow 1 , b \rightarrow 0$) will be written as
\bea\label{tz} 
g<1 \hspace{0.5cm}&;&\hspace{0.5cm} 
T^{\dagger} \z{1(3)} T = s'^z  \;\hspace{0.5cm};\hspace{0.5cm}
T^{\dagger} \z{2} T = -s'^z    \\ \nonumber
g>1 \hspace{0.5cm}&;&\hspace{0.5cm} 
T^{\dagger} \z{1(3)} T = \frac{-I}{2} \hspace{0.5cm};\hspace{0.5cm}
T^{\dagger} \z{2} T = s'^z .   
\eea
The magnetization ($m=\frac{1}{N} \sum_{i}^{N} \bra{0}\z{i}\ket{0}$) is
\bea
m&=&\frac{1}{N} \sum_{\mu=1}^{N/3} 
\bra{0}(\z{1\mu}+\z{2\mu}+\z{3\mu})\ket{0}  \\ \nonumber
&=&\frac{1}{N} \sum_{\mu=1}^{N/3}
\bra{'0}T^{\dagger}(\z{1\mu}+\z{2\mu}+\z{3\mu})T\ket{0'}
\eea
where $\ket{0'}$ is the ground state in the renormalized Hilbert space 
($T\ket{0'}=\ket{0}$). By using equation (\ref{tz}), the magnetization is 
calculated to be 
\bea\label{mmp}
m=\frac{1}{3}m'  \hspace{0.5cm}&;&\hspace{0.5cm} g<1 \\ \nonumber
m=\frac{-1}{3}+\frac{1}{3}m' \hspace{0.5cm}&;&\hspace{0.5cm} g>1 
\eea
where $m'$ is the magnetization in the renormalized chain with $\frac{N}{3}$
sites. However in the thermodynamic limit ($N\rightarrow\infty$), $m$ and $m'$
will be equal. Thus equation (\ref{mmp}) gives
\bea\label{mag}
m=0 \hspace{0.5cm}&;&\hspace{0.5cm} g<1 \\ \nonumber
m=-0.5 \hspace{0.5cm}&;&\hspace{0.5cm} g>1 .
\eea
Similarly, we can use the definition of staggered magnetization
($sm=\frac{1}{N} \sum_{i}^{N} \bra{0}(-1)^i \z{i}\ket{0}$) and repeat the steps 
in calculating $m$, the staggered magnetization is obtained to be 
\bea\label{smag}
sm=0.5 \hspace{0.5cm}&;&\hspace{0.5cm} g<1  \\ \nonumber
sm=0 \hspace{0.5cm}&;&\hspace{0.5cm} g>1.
\eea
These results confirm the Neel ordered state in the $g<1$ region and the 
saturated ferromagnetic state in the $g>1$. Note that these results are
the same as the exact ones which have been obtained in the last section.
Had we taken even size blocks for the renormalization procedure, we could not 
obtain these values.

The calculation of the ground state energy is easily done by accumulating the 
energy of blocks in a hierarchical way. The renormalized Hamiltonian is 
\be
H'_{\frac{N}{3}}(k',g')=T^{\dagger} H_N(k,g) T=\left\{ \begin{array}{cc}
\frac{k}{4}(\frac{-2N}{3})+
k \sum_{i}^{N/3}(\z{i}'\z{i+1}'+g\z{i}') \;\;\;;\;\;\; g<1 \\
\frac{kN(1-4g)}{12}+k(g-1) \sum_{i}^{N/3}\z{i}' \;\;\;\;\;\;\;\;\;\;\;;\;\;\; g>1\\
\end{array}\right.
\ee
Therefore the ground state energy per site is calculated to be
\be
e_o=\frac{E_o}{kN}=\left\{ \begin{array}{cc}
\frac{1}{4}(\frac{-2}{3})(1+\frac{1}{3}+\frac{1}{9}+ ... )=
\frac{-1}{4} \;\;\;;\;\;\; g<1 \\
\frac{1-4g}{12}-\frac{2(g-1)}{12}=\frac{1-2g}{4} 
\;\;\;\;\;\;\;\;\;\;\;\;\;\;;\;\;\; g>1\\
\end{array}\right.
\ee
which is equal to the exact ground state energy. This result can not  
be either obtained 
by taking an even size block in the RG procedure.

\newpage
{\Large \bf Tables}
\vspace{0.5cm}

Table-1. Eigenvalues and critical exponents at the XXTF fixed point, 
both RSRG and exact results.
$$
\begin{array}{|c|c|c|c|c|c|c|}\hline
&\lambda_1&\lambda_2&\beta&\nu&z&\alpha\\ \hline
RSRG&0.250&4.000&0.792&0.792&1.262&0.208 \\ \hline
Exact&-&-&0.5&0.5&2&0.5   \\ \hline
\end{array}
$$

\newpage
{\Large \bf Figure Captions :}
\vspace{0.5cm}

{\bf Fig.1)} Decomposition of lattice into block and inter--block part,
and different types of intra--block ($H^B$) and inter--block ($H^{BB}$) interactions.

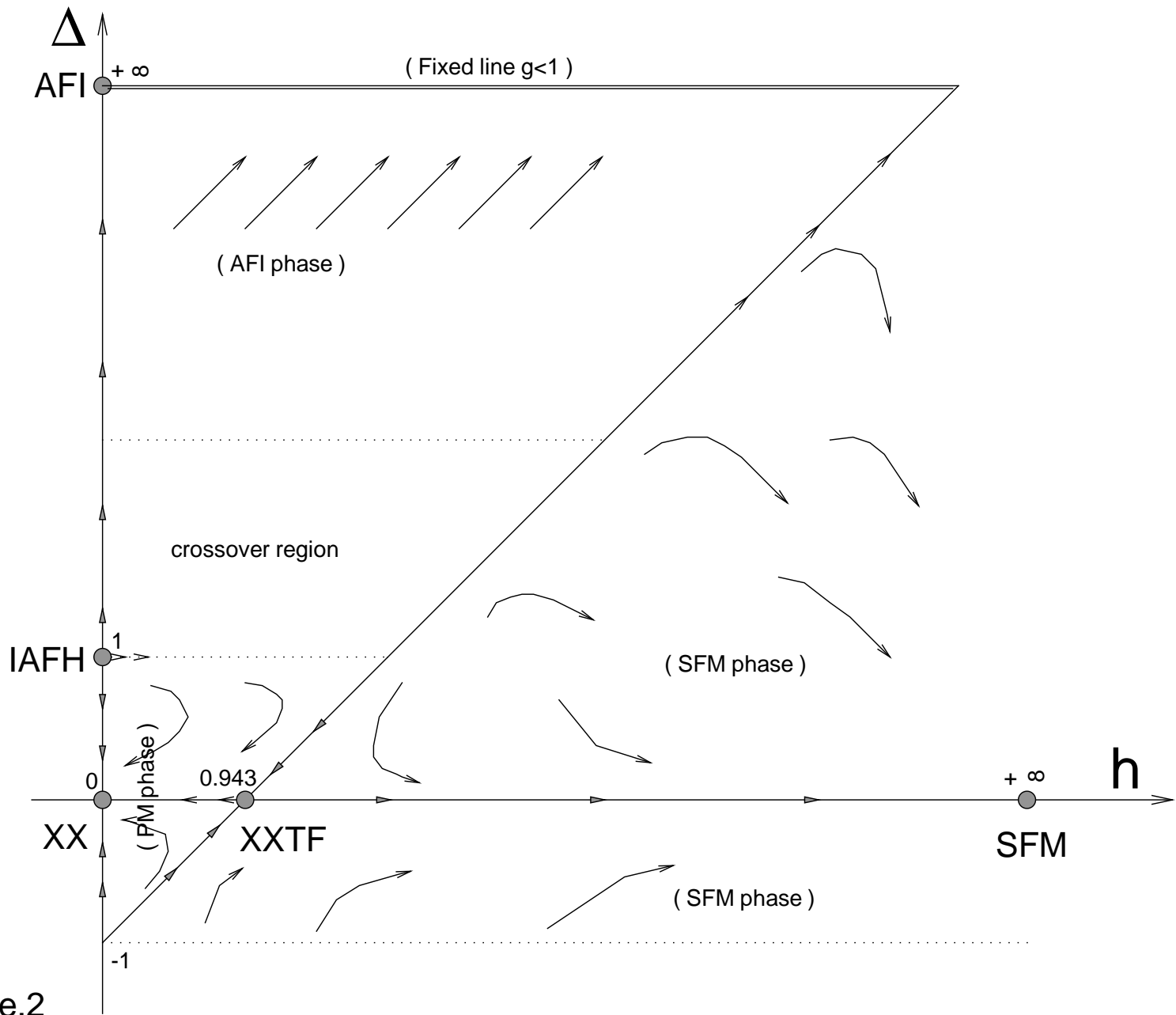
{\bf Fig.2)} Phase diagram of the {\bf xxz} model in the presence of 
an external magnetic field ($h$). Filled circles are the fixed points and
arrows show the direction of flow. The solid line which passes through 
the ($\d=-1, h=0$)
and($\d=1, h=1.886$) points is the critical line $h_c=0.943\d+0.943$.
The dotted line for 
$\d>1$ show qualitatively the crossover region. The double solid line 
at $\d=\infty$ is the fixed line $0<g\leq1$.

{\bf Fig.3-a)} Ground state energy per site versus external field ($h$) for
$\d=0$ , both RSRG and exact results.

{\bf Fig.3-b)} Ground state energy per site versus external field ($h$) for
$\d=-1, -0.5, 0.5, 1$, which has been obtained by RSRG.

{\bf Fig.4-a)} Magnetization ($m=\frac{1}{N}\sum_{i=1}^N \bra{0}\z{i}\ket{0}$) versus
external field ($h$) for $\d$=0, both RSRG and exact results.

{\bf Fig.4-b)} Magnetization ($m=\frac{1}{N}\sum_{i=1}^N \bra{0}\z{i}\ket{0}$) versus
external field ($h$) for $\d=-0.5, 0.5$, which has been obtained by RSRG.

{\bf Fig.4-c)} Magnetization ($m=\frac{1}{N}\sum_{i=1}^N \bra{0}\z{i}\ket{0}$) versus
external field ($h$) for $\d=1.1, 1.5$, which has been obtained by RSRG.

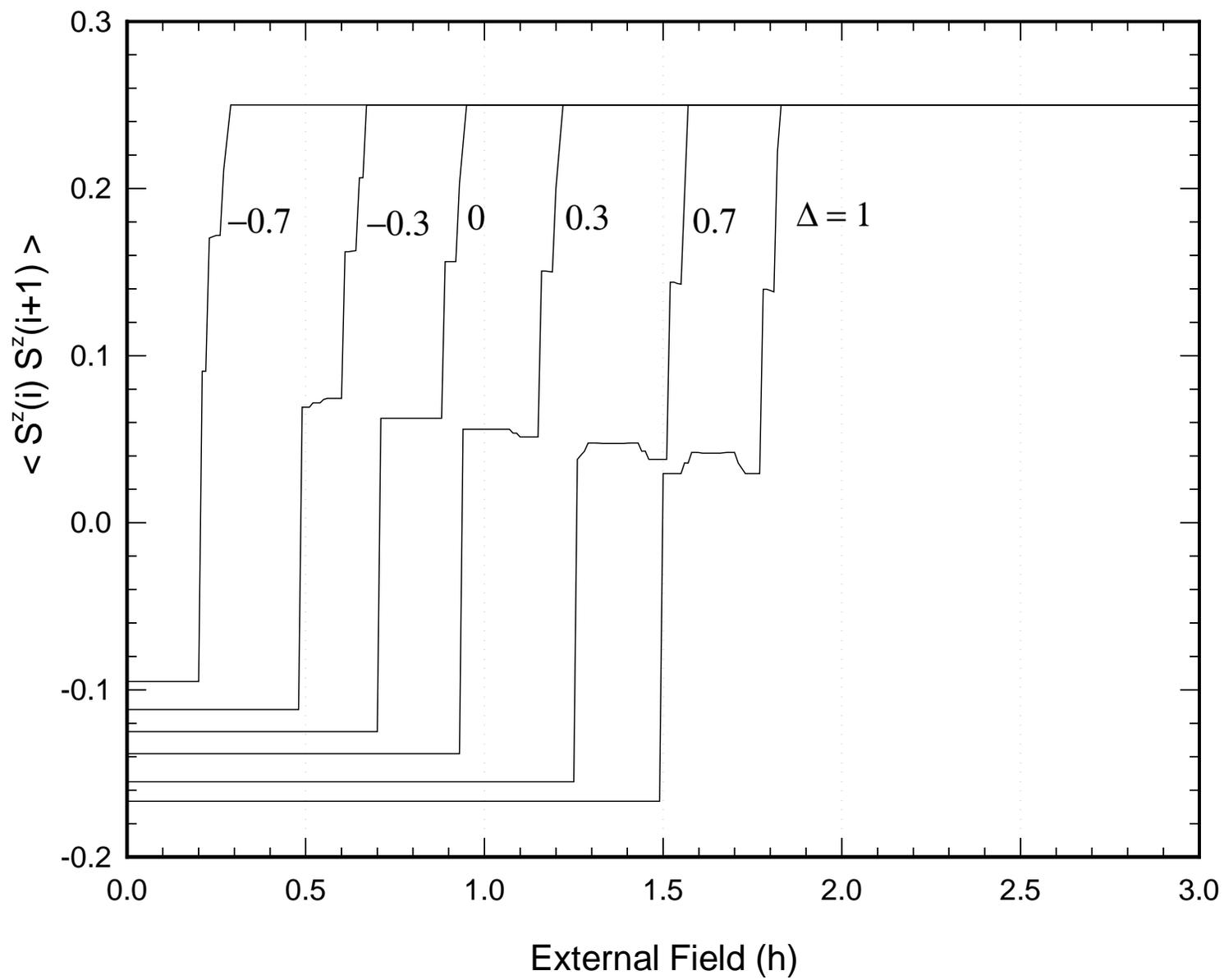
{\bf Fig.5)} z--component of spin--spin correlation function for different
value of anisotropy parameter($\d$) versus external field ($h$).

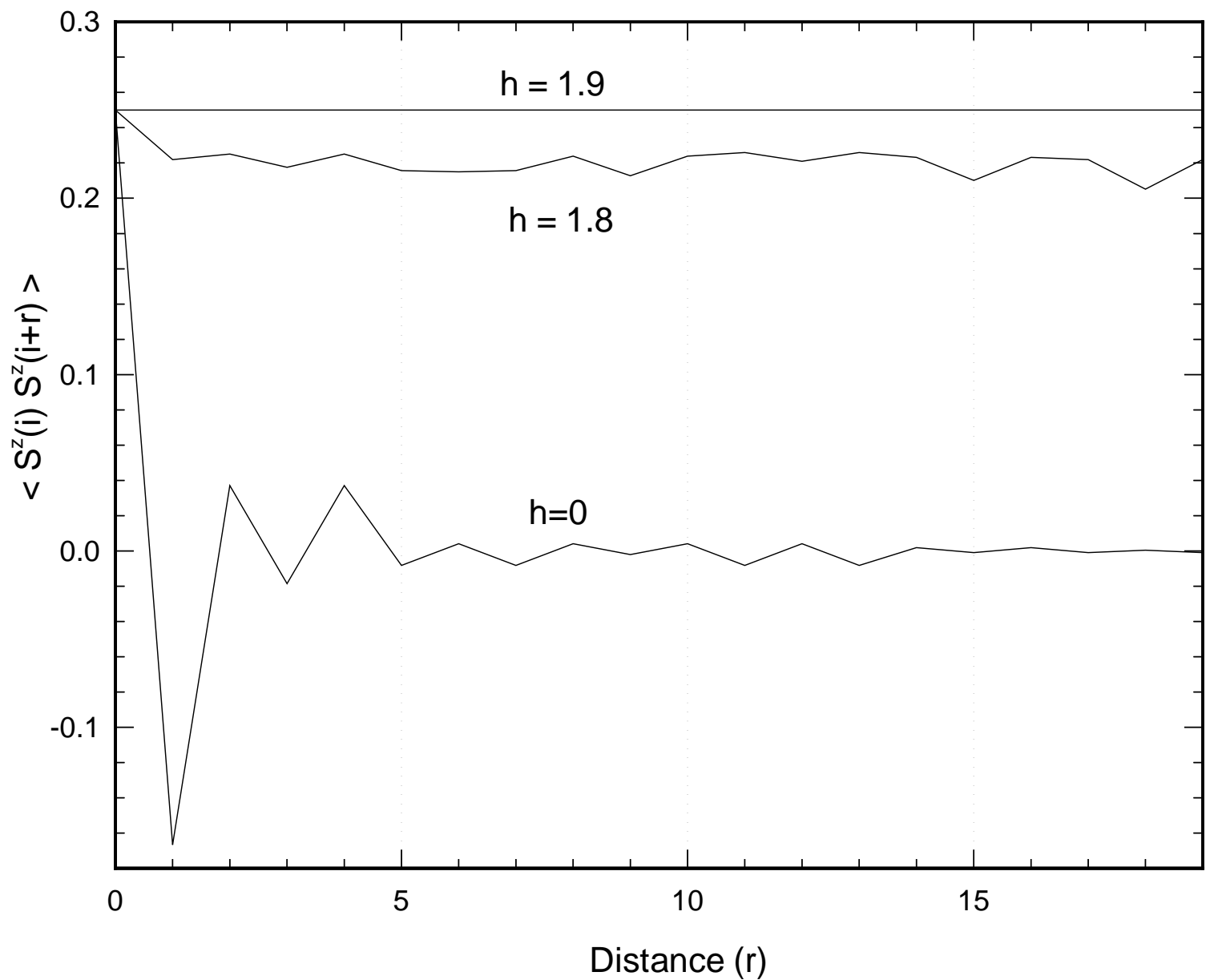
{\bf Fig.6)} z--component of spin--spin correlation function at $\d=1$
versus distance ({\bf r}) for different value of external field ($h$).


\newpage
\begin{thebibliography}{99}
\bibitem{r1} K. G. Wilson, Rev. Mod. phys. {\bf 47}, 773 (1975).
\bibitem{r2} S. D. Drell, M. Weinstein and S. Yankielowicz, Phys. Rev. 
{\bf D14}, 487 (1976).
\bibitem{r3} P. Pfeuty, R. Jullien and K. L. Penson in: Real-Space Renormalisation,
eds. T. W. Burkhardt and J.M.J. van Leeuwen (Springer, Berlin, 1982) ch.5;
G. Sierra and M.A. Martin-Delgado in : Strongly Correlated Magnetic 
and Superconducting Systems, Lecture notes in physics Vol. 478, 
(Springer, Berlin, 1997).
\bibitem{r4} A. L. Stella, C. Vanderzande and R. Dekeyser, Phys. Rev.
{\bf B27}, 1812 (1983).
\bibitem{r5} M.A. Martin-Delgado and G. Sierra,
Int. J. Mod. Phys. {\bf A11}, 3145 (1996);
J. Gonzalez, M.A. Martin-Delgado, G. Sierra, A. H. Vozmediano in :
Quantum Electron Liquids and High-$T_{c}$ Superconductivity, Lecture Notes
in Physics, Monographs. Vol 38 (Springer, Berlin, 1995) ch.11.
\bibitem{r6} S. R. White and R. M. Noak, Phys. Rev. Lett. {\bf 68}, 3487 (1992).
\bibitem{r7} M.A. Martin-Delgado and G. Sierra, Phys. Lett. {\bf B364}, 41 (1995),
M.A. Martin-Delgado, J. Rodriguez-Laguna and G. Sierra, 
Nucl. Phys. {\bf B473}, 685 (1996).
\bibitem{r8} A. Langari and V. Karimipour, Phys. Lett {\bf A236}, 106 (1997).
\bibitem{r9} S. R. White, Phys. Rev. Lett. {\bf 69}, 2863 (1992); S. R. White,
Phys. Rev. {\bf B48}, 10345 (1993).
\bibitem{r10} A. Drzewinski and J.M.J. van Leeuwen, Phys. Rev. {\bf B49}, 403 (1994).
\bibitem{r11} A. Drzewinski and R.Dekeyser, Phys. Rev. {\bf B51}, 15218 (1995).
\bibitem{r12} C. N. Yang and C. P. Yang, Phys. Rev. {\bf 150}, 321 (1966).
\bibitem{r13} H. Bethe, Z. Physik {\bf 71}, 205 (1931); R. Orbach, phys. Rev.
{\bf 112}, 309 (1958); J. D. Cloizeaux and M. Gaudin, J. Math. Phys. {\bf 7}, 1384 (1966).
\bibitem{r135} V. E. Korepin, A. G. Izergin and N. M. Bogoliubov in :
Quantum Inverse Scattering Method, Correlation Functions and Algebric
Bethe Ansatz (Cambridge : Cambridge University Press 1993).
\bibitem{r14} F. C. Alcaraz and A. L. Malvezzi, J. Phys. A. Math. Gen. 
{\bf 28}, 1521 (1995); F. C. Alcaraz, M. N. Barber and M. T. Batchelor,
Phys. Rev. Lett. {\bf 58}, 771 (1987); Ann. Phys. {\bf 182}, 280 (1988);
C. J. Hammer, J. Phys. A. Math. Gen. {\bf 19}, 3335 (1986);
H. J. de Vega and F. Woynarovich, Nucl. Phys. {\bf B251}, 439 (1985);
F. Woynarovich, Phys. Rev. Lett. {\bf 59}, 259 (1987).
\bibitem{r15} J. M. Rabin, Phys. Rev. {\bf B21}, 2027 (1980).
\bibitem{r16} E. H. Lieb, T. Schultz and D. Mattis, Ann. Phys. {\bf 16}, 407 (1961);
E. Barouch and B. M. McCoy, Phys. Rev. {\bf A3}, 786 (1971);
T. Niemeijer, Physica, {\bf 36}, 377 (1967);
P. Pfeuty, Ann. Phys. {\bf 57}, 79 (1970).
\bibitem{r165} V. Pasquier and H. Saluer, Nucl. Phys. {\bf B330}, 523 (1990);
and references therein.
\bibitem{r17} M. P. A. Fisher, P. B. Weichman, G. Grinstein and D. S. Fisher,
Phys. Rev. {\bf B40}, 546 (1989).
\bibitem{r18} M.A. Martin-Delgado and G. Sierra, Phys. Rev. Lett. 
{\bf 76}, 1146 (1996).

\end{thebibliography}
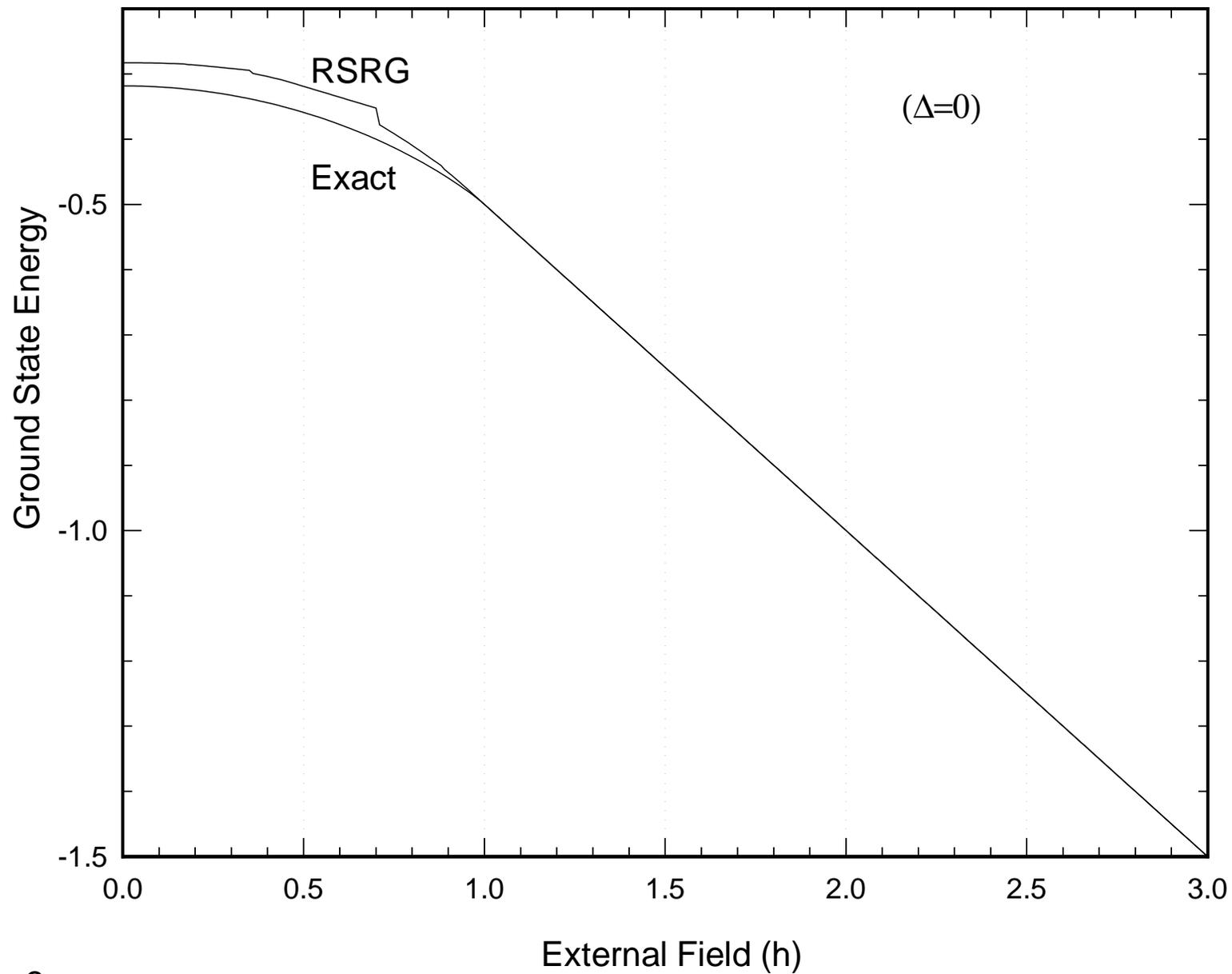
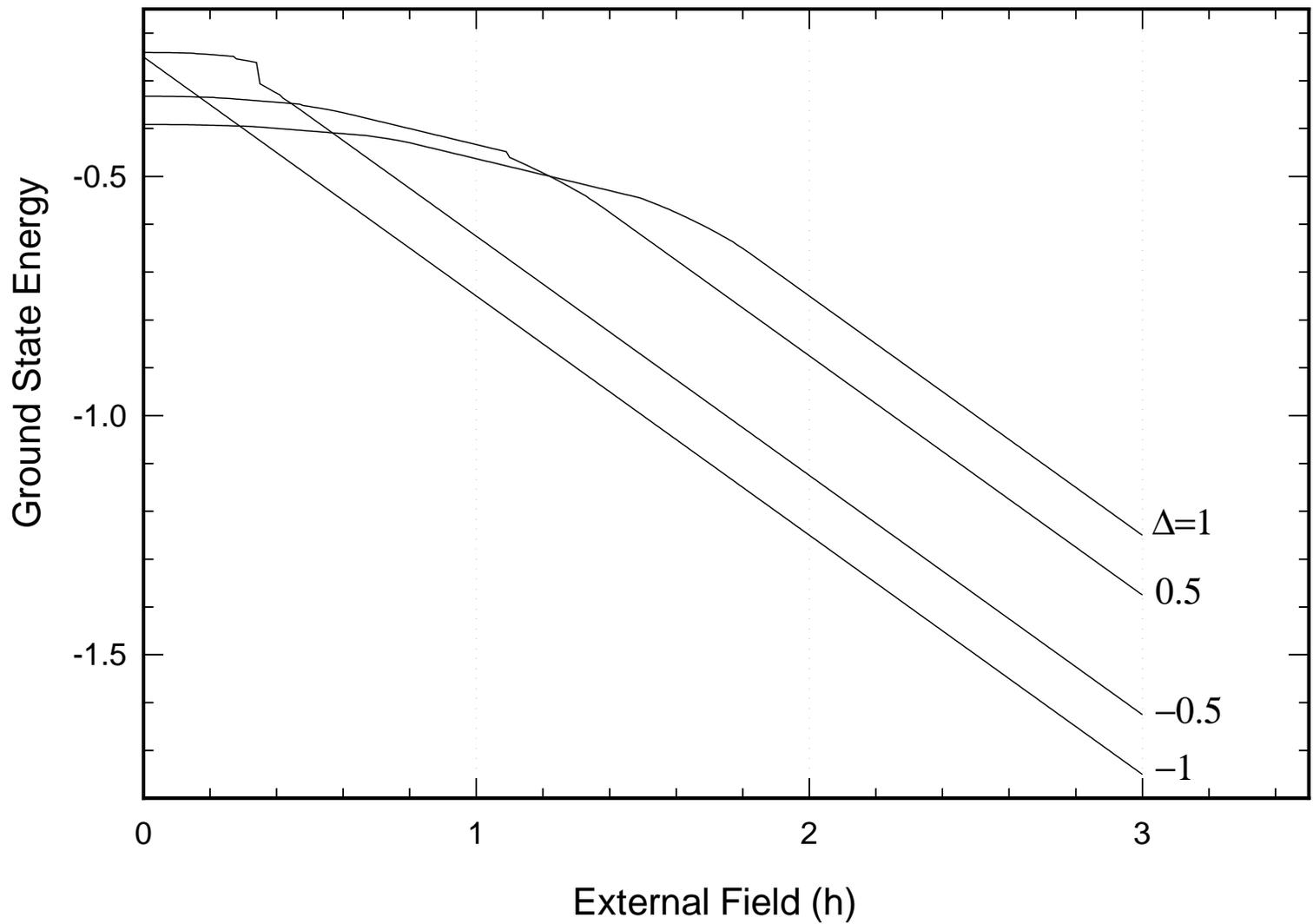
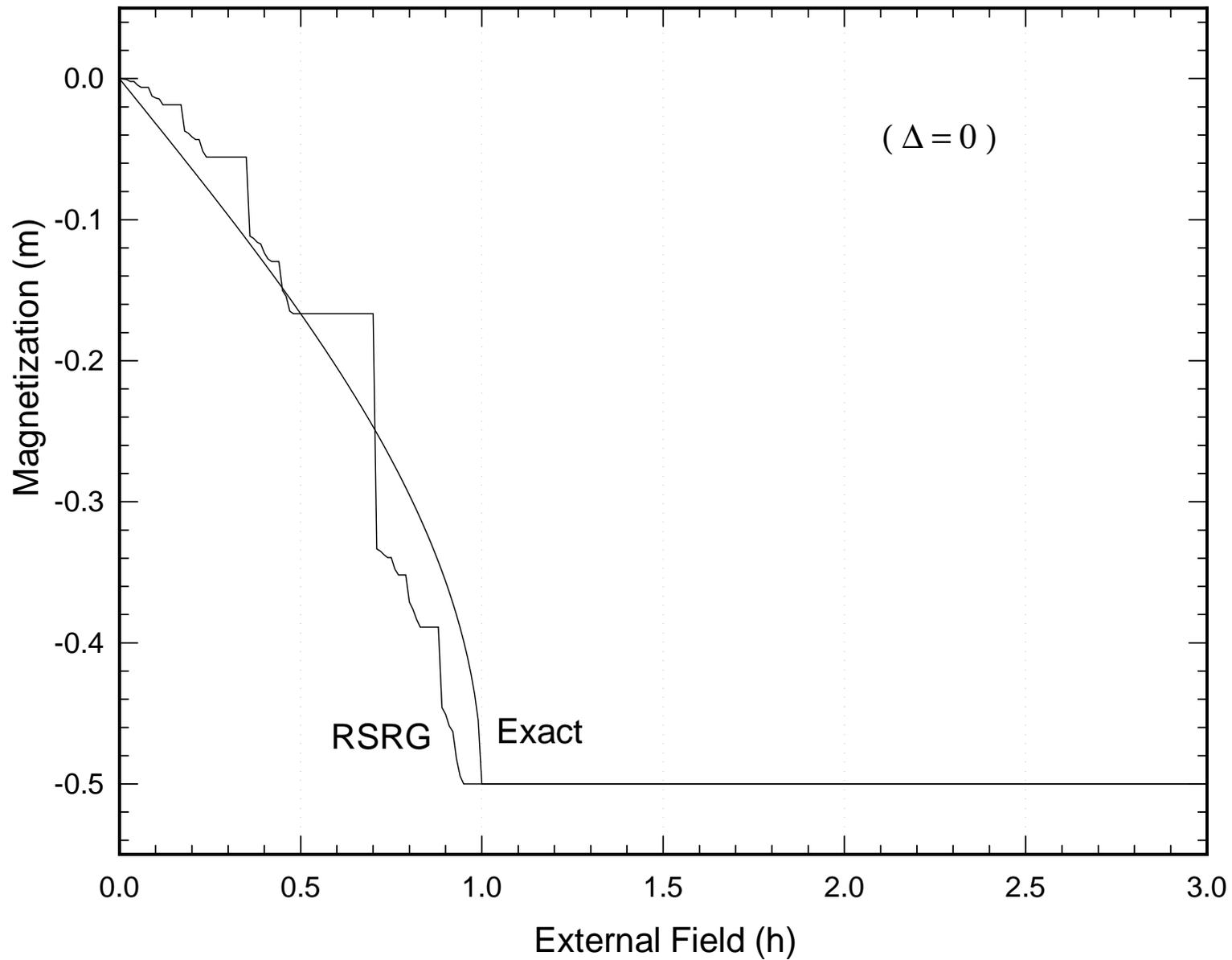
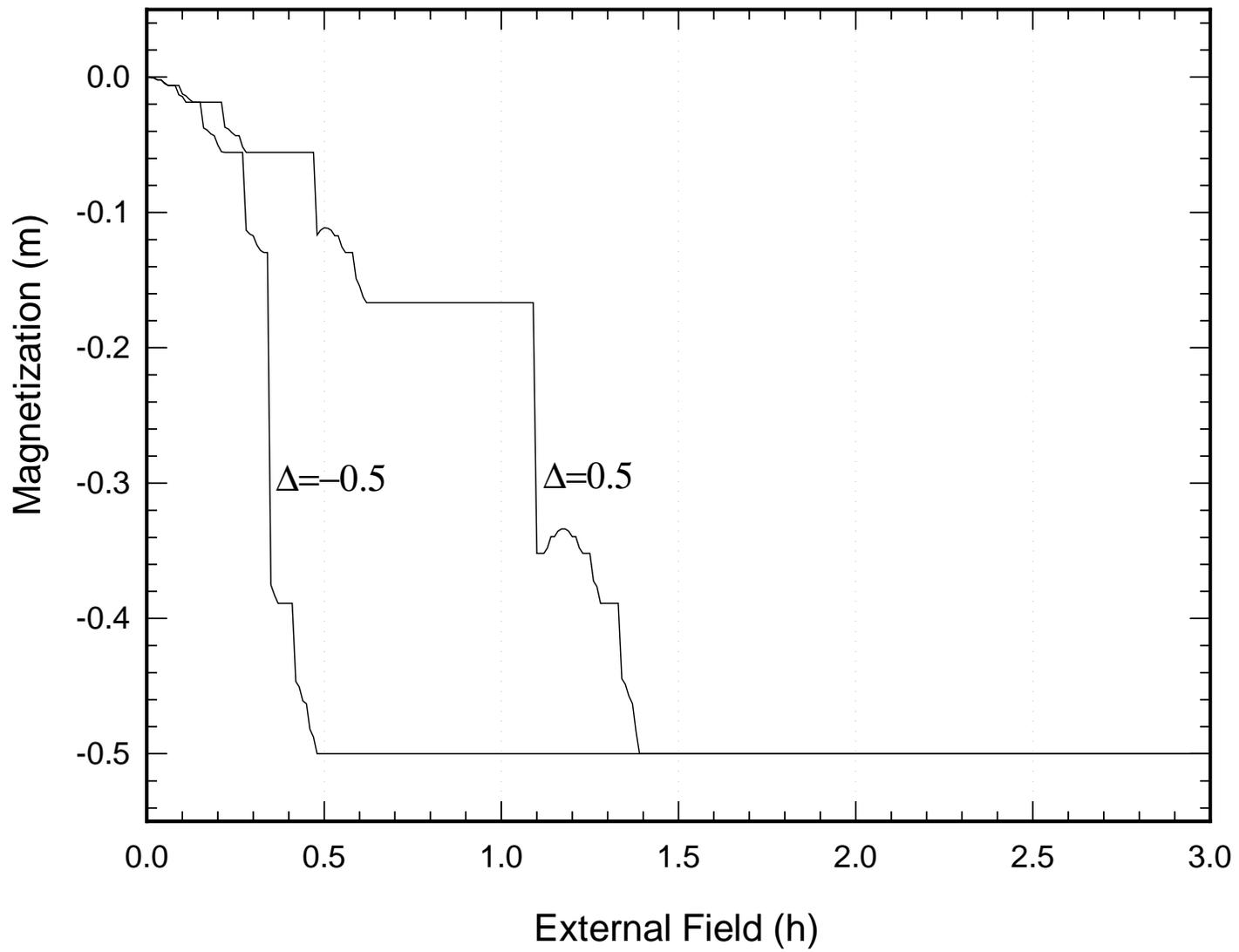
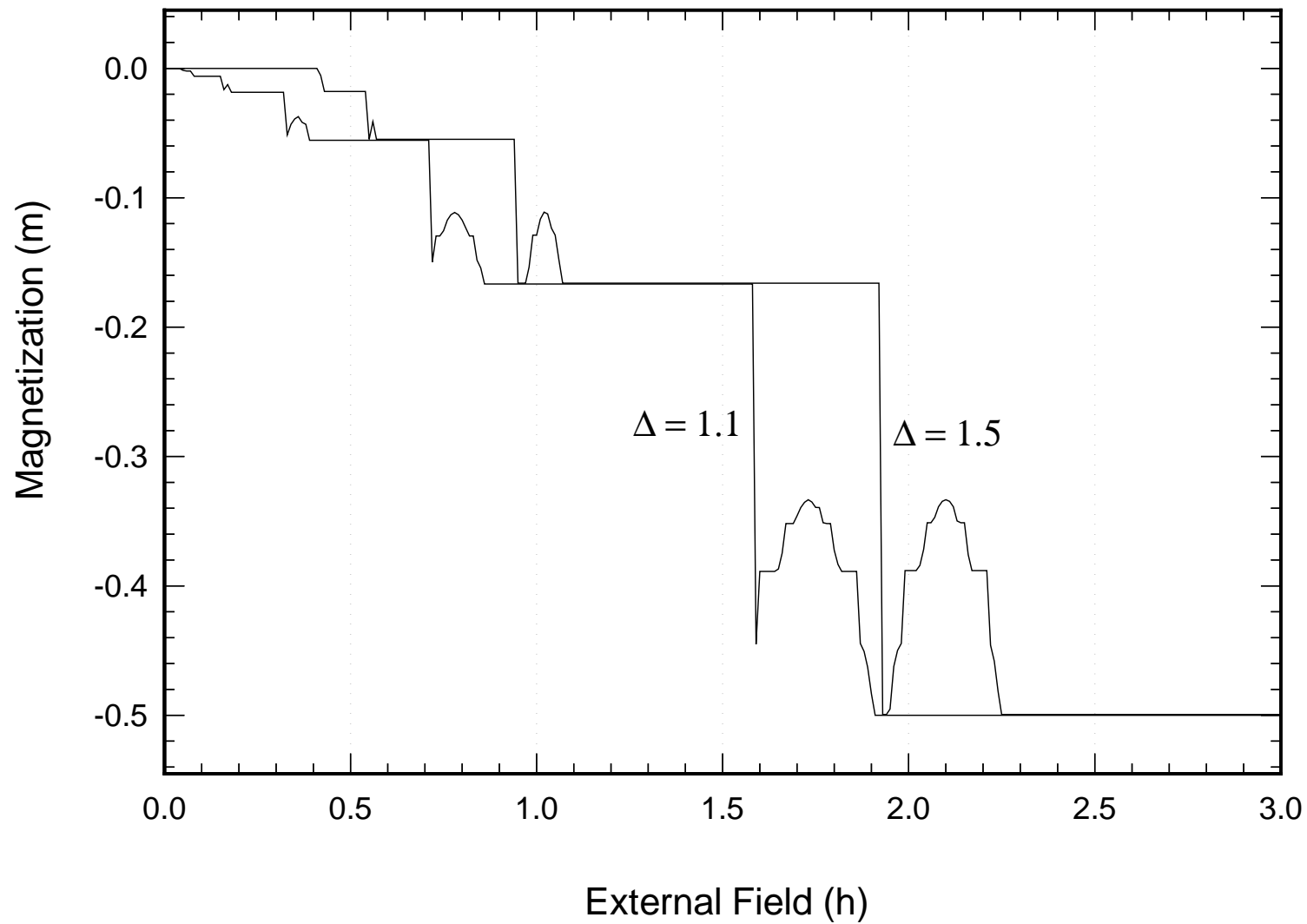
\end{document}